\documentclass[nolinenumber]{aastex631}

\newcommand{\tianshu}[1]{{#1}}

\usepackage{amsmath}

\begin{document}

\title{Insights into the Production of $^{44}$Ti and Nickel Isotopes in Core-Collapse Supernovae}

\author[0000-0002-0042-9873]{Tianshu Wang}
\affiliation{Department of Astrophysical Sciences, Princeton University, Princeton, N.J. 08544}

\correspondingauthor{Tianshu Wang}
\email{tianshuw@princeton.edu}

\author[0000-0002-3099-5024]{Adam Burrows}
\affiliation{Department of Astrophysical Sciences, Princeton University, Princeton, N.J. 08544}
\affiliation{Institute for Advanced Study, 1 Einstein Drive, Princeton, NJ 08540}





\begin{abstract}
We report nucleosynthetic results for both $^{44}$Ti and nickel isotopes for eighteen three-dimensional (3D) core-collapse supernova (CCSN) simulations extended to $\sim$20 seconds after bounce. We find that many of our long-term models are able to achieve $^{44}$Ti/$^{56}$Ni ratios similar to that observed in Cassiopeia A, and modern supernova models can synthesize up to $2\times10^{-4}M_\odot$ of $^{44}$Ti. Neutrino-driven winds and the fact that there can be simultaneous accretion and explosion in 3D models of core-collapse supernovae play central roles in its production. We conclude that the $^{44}$Ti underproduction problem in previous CCSN models is no longer an issue. In addition, we discuss the production of both $^{57}$Ni and stable nickel/iron ratios and compare our results to observations of SN1987A and the Crab. 


\end{abstract}

\keywords{Supernova, Nucleosynthesis}


\section{Introduction}
Supernova remnants offer unique laboratories for the study of core-collapse supernova (CCSN) explosions. Various aspects of the CCSN theory can be constrained by measuring the abundance and distribution of different nuclei in the remnants. Despite of the fact that CCSNe are able to produce many of the elements in the periodic table \citep{arcones2023}, there are a few isotopes (such as the radioactive isotopes $^{44}$Ti, $^{56}$Ni, and $^{57}$Ni) that are of particular interest both theoretically and observationally. These isotopes are created in the core of the supernova explosion, and as a result their formation is closely related to the details of the explosion mechanism itself. Two of them play important roles in the evolution of supernovae light curves because they radioactively heat the ejecta. Furthermore, with relatively short half-lives such isotopes are ideal targets for $\gamma-$ray astronomy \citep{kurfess1992,iyudin1994,iyudin1998,diehl1998,renaud2006,boggs2015,weinberger2020}. In addition, supernovae produce interesting stable nickel and iron isotopes, either directly or after decay. 
Hence, understanding the production and distribution of such isotopes in core-collapse explosions offers us a direct connection between theory and observation.  Given this, it is important to make reliable theoretical predictions of supernova nucleosynthetic yields. 

It is well known that multi-dimensional effects play an important role in CCSN explosions (for reviews, see \citet{janka2012,burrows2021,wang2022}), and such effects can significantly influence nucleosynthetic results. For example, the high $^{44}$Ti masses (at least $10^{-4} M_\odot$) observed in SN1987A and Cassiopeia A (Cas A) are strongly inconsistent with spherically-symmetric predictions \citep{the2006,sukhbold2016,curtis2019,sieverding2023}. To correctly capture the role of multi-dimensional effects and to make reliable theoretical predictions of CCSN ejecta abundances, high-fidelity three-dimensional state-of-the-art CCSN simulations are required. However, due to computational resource limitations, previous 3D CCSN models \citep{wongwathanarat2017,muller2017,stockinger2020,bollig2021,sandoval2021,sieverding2023} that generated nucleosynthetic yields had at least one of the following unsatisfactory aspects: (a) the neutrino-transport scheme was approximate and the ejecta electron fractions ($Y_e$) were very uncertain; (b) the radiation transport calculation (or the simulation itself) was turned off long before reaching the asymptotic stage of explosion and the contribution of neutrino-driven winds was ignored; (c) the calculation of multi-dimensional hydro/radiation effects was switched to sphericized and/or parameterized models after only one or two seconds post-bounce when asymmetric long-lasting accretion is still strong; and (d) the size of the nucleosynthesis network was too small to reliably estimate the production of many isotopes. Most importantly, since previous work focused on a small number of models, the robustness and the progenitor dependencies of the results remained unknown. 

In this work, we report the nucleosynthetic results of 18 3D CCSN simulations using the code F{\sc{ornax}} \citep{skinner2019}. The simulations were evolved to late times after bounce (many seconds, see Table \ref{tab:simulation_summary}) with full neutrino transport and multi-dimensional effects, and the asymptotic explosion energies have been reached for most models. By extrapolating both the tracer thermal histories and the neutrino-driven wind mass outflow rates, we are able to calculate the ejecta isotopic abundances to about twenty seconds post-bounce. Hence, for the first time one is able to perform a systematic study of the asymptotic nucleosynthetic yields in 3D CCSN models, including the contribution of neutrino-driven winds.

The paper is arranged as follows: We summarize the simulation specifics and the approach to the nucleosynthetic calculations in Section \ref{sec:method}.  In this section we also describe the extrapolation methods adopted. The main results are presented in Section \ref{sec:results}, in which we provide a detailed discussion of the production of the radioactive isotopes $^{44}$Ti, $^{56}$Ni, and $^{57}$Ni. We find that all of our 3D models are able to reach a  $^{44}$Ti/$^{56}$Ni ratio similar to that of Cas A, and we find that many models can produce more than $10^{-4}M_\odot$ of $^{44}$Ti. Therefore, the high $^{44}$Ti problem \citep{the2006} of the past is naturally solved by long-term self-consistent 3D CCSN models, without the need to introduce fast initial rotation or jet-like structures. We find that the $^{57}$Ni/$^{56}$Ni ratio of SN1987A (though with large errorbars) is only about 1.5 times higher than our models. Since $^{57}$Ni is a neutron-rich isotope, it is possible that SN1987A contains slightly more neutron-rich ejecta than expected, which might arise from a slightly more prompt explosion than our models would suggest. We study the influence of different extrapolation methods in Section \ref{sec:compare} and we also briefly discuss the stable Ni/Fe ratios we find.
Finally, in Section \ref{sec:conclusion} we summarize the paper and discuss the uncertainties and caveats of our work.

\section{Method}\label{sec:method}
In this paper, we provide a subset of the nucleosynthetic results for eighteen CCSN models with solar-metallicity progenitors sporting ZAMS masses that range from 9 to 60 $M_\odot$. Two models (the 19.56 and 40 $M_{\odot}$) form black holes and experience successful explosions \citep{burrows2023,burrows2024}. All these models have already been published in \citet{wang2024} and \citet{burrows2024}. Table \ref{tab:simulation_summary} summarizes some basic properties of the models. Since the publication of \citet{wang2024} and \citet{burrows2024}, some of the models (the 17, 18, 18.5, 20, 25, 60 $M_\odot$ models) have been simulated to even later times after bounce and these updates are reflected in what we present here.

All models are generated using the multi-group radiation/hydrodynamics code F{\sc{ornax}} \citep{skinner2019,vartanyan2019,burrows2019,burrows2020} with 12 logarithmically-distributed energy groups for each of our three neutrino species (electron-type, anti-electron type, and ``$\mu$-type," a bundling of the $\mu$ and $\tau$ neutrinos and anti-neutrinos). The SFHo nuclear equation of state (EOS) of \citet{steiner2013} is used. Each simulation is done with a spherical grid with (1024, 128, 256) cells along the $r$, $\theta$, and $\phi$ directions, and the outer boundary radii vary from 30,000 to 100,000 kilometers (km). Each model is first simulated spherically until 10 milliseconds after bounce, after which it is continued in full 3D. No perturbations are introduced when the 3D simulations are started, except in the 9a model. The initial perturbations in 9a model cause a slightly more prompt explosion with more neutron-rich ejecta, which is very different from the perturbation-free 9b model of the same progenitor \citep{wang2024}. However, as shown in \citet{wang2024}, realistic initial perturbations in more massive models don't seem to influence the electron fraction ($Y_e$) or nucleosynthetic yields of the ejecta.

Nucleosynthetic yields in this paper are calculated in the same way as in \citet{wang2024}. We use post-processed backwardly-integrated tracer particles to track the dynamical and thermal evolution of the ejected matter. About 320,000 tracers are injected into each simulation. The tracers are placed logarithmically along the r-direction above 500 km and uniformly along the $\theta$- and $\phi$-directions at the end of each simulation. A more detailed description of the tracer method can be found in \citet{wang2024}. The tracer trajectories with neutrino field information are fed into SkyNet \citep{lippuner2017}, and a 1530-isotope network including elements up to $A=100$ is used to calculate the nucleosynthetic results. The reaction rates are taken from the JINA Reaclib \citep{cyburt2010} database, and we include neutrino interactions with protons and neutrons. However, reactions for the $\nu-$process \citep{woosley1990} are not included. The NSE (nuclear statistical equilibrium) criterion is set at 0.6 MeV ($\sim$7 GK). The electron fractions ($Y_e$) are calculated by F{\sc{ornax}} when the temperature is above the NSE threshold, which allows the neutrino spectra to be appropriately non-thermal. A nucleosynthesis calculation starts from the point after which the tracers never reach NSE again. We use the \citet{lodders2021} values whenever solar abundances are referred to in this paper. \tianshu{If not otherwise stated, all ratios mentioned in the paper are mass fraction ratios.}

Although our 3D CCSN models have been evolved to many seconds post-bounce (see Table \ref{tab:simulation_summary}), we find that some of them have not been carried long enough to reach the asymptotic abundances of certain isotopes. The first reason for this is that not all the tracers representing the ejecta have cooled down and finished their nucleosynthetic evolution, especially those ejected in the neutrino-driven wind phase. The second reason is that even at the end of our longest simulations, the neutrino-driven wind is still actively producing certain isotopes at a non-negligible rate. Therefore, to get the final ejecta isotopic abundance, two kinds of extrapolations are required: (a) extrapolation of the tracer thermal histories to calculate the asymptotic abundances of the matter ejected before the end of our simulations and (b) extrapolation of the mass outflow rates of the neutrino-driven winds to estimate how much more matter will be ejected after the end of our 3D simulations.

As a result, we extrapolate the tracer thermal histories of the 3D simulations using a power-law method \citep{ning2007,zha2024}. The temperatures and densities at the end of our simulations are continued using
\begin{equation}
    \begin{split}
        T(t) =& T(t_{\rm end})(1+(t-t_{\rm end})/\tau)^{-2/3},\\ 
        \rho(t) =& \rho(t_{\rm end})(1+(t-t_{\rm end})/\tau)^{-2},\\ 
    \end{split}
\end{equation}
where $t_{\rm end}$ is the end time of each simulation, and $\tau$ is the expansion timescale fitted from the last 10 ms of the associated trajectories of the 3D simulations. There is also an exponential method used in the literature for trajectory extrapolations \citep{magkotsios2011}. The impact on nucleosynthetic results caused by different trajectory extrapolation methods will be discussed in more detail in Section \ref{sec:compare}. 

\begin{figure}
    \centering
    \includegraphics[width=0.96\textwidth]{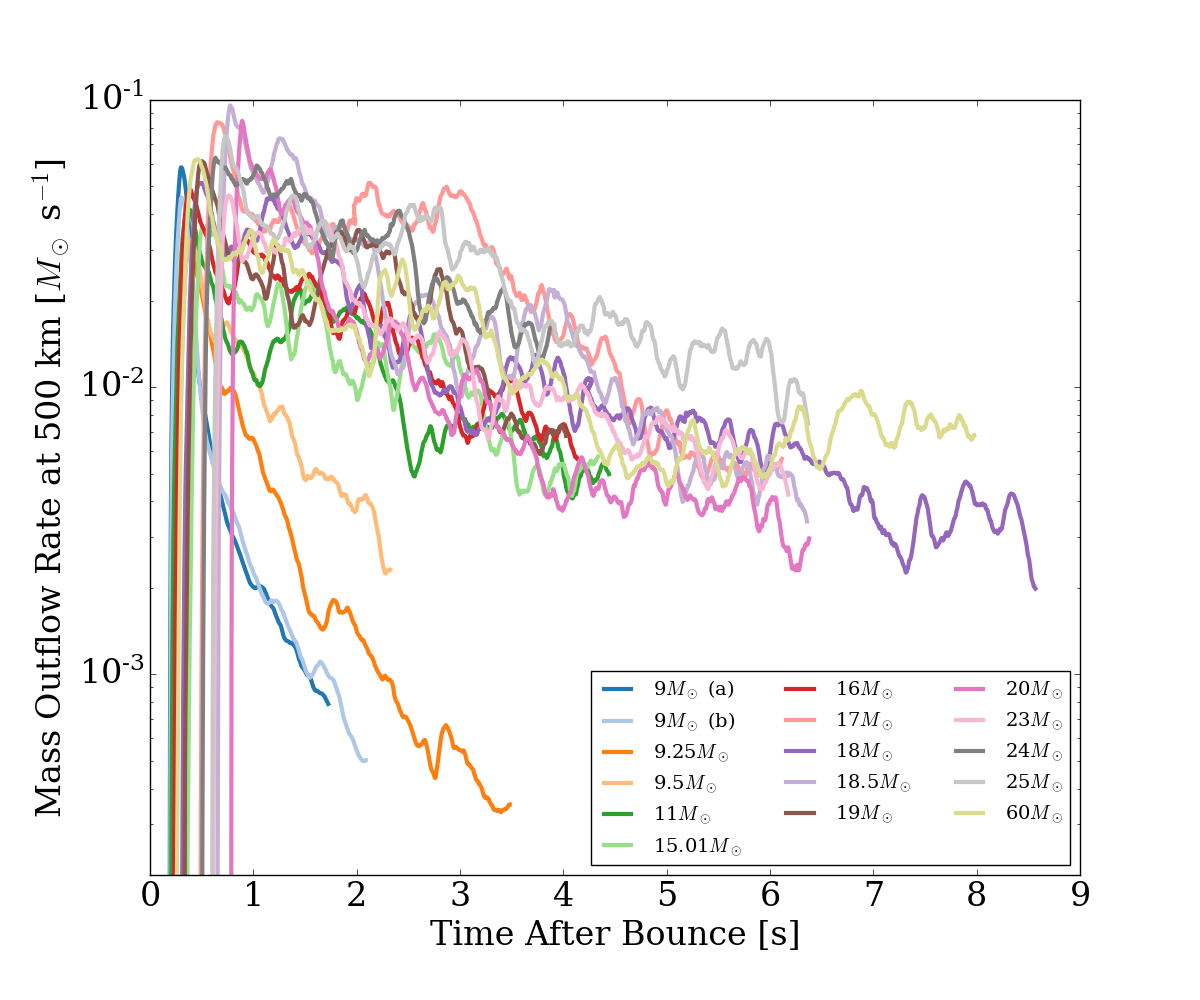}
    \caption{Neutrino-driven wind mass outflow rates measured at 500 km, smoothed over 100 milliseconds. Although there are fluctuations in the wind mass outflow rates, they generally decay exponentially with time at later times, which allows us to extrapolate their evolution. We fit exponential relations to the wind mass outflow rates in the last 1.5$-$2 seconds of each simulation. The black-hole formers (19.56 and 40 $M_{\odot}$) are not shown on this plot because their neutrino-driven winds are instantly turned off after the black hole formation (which is the end of the 3D simulation).}
    \label{fig:wind}
\end{figure}

Figure \ref{fig:wind} shows the mass outflow rates of the neutrino-driven winds measured at 500 km as a function of time for all our models, smoothed over 100 milliseconds. We fit an exponential relation to the last one second of each simulation. We choose the exponential form not only because it fits the data well, but also to make sure that the total amount of mass ejected in the extrapolated phase is always convergent. We assume that the future ejected matter will have the same asymptotic abundances as the average values of tracers ejected in the last second. The black-hole formers don't need this type of extrapolation because their neutrino-driven winds are instantly turned off after black hole formation (which is at the end of the 3D simulation).

With the extrapolation methods described above, we calculate the nucleosynthetic evolution of all models to about twenty seconds post-bounce. This allows us to study the final yields of the important isotopes we discuss in this work.

\begin{table}
    \centering
    \begin{tabular}{c|ccccccc}
    $M_{\rm ZAMS}$  &$t_{\rm end}$  &$E_{\rm exp}$  &\tianshu{$M^{\rm pow}_{^{44}{\rm Ti}}$($M^{\rm exp}_{^{44}{\rm Ti}}$)}  &\tianshu{$M^{\rm pow}_{^{56}{\rm Ni}}$($M^{\rm exp}_{^{56}{\rm Ni}}$)}  &\tianshu{$M^{\rm pow}_{^{57}{\rm Ni}}$($M^{\rm exp}_{^{57}{\rm Ni}}$)}  &Stable Ni/Fe &\tianshu{$M_{\rm extra}$}\\
    $[M_\odot]$     &[s]            &[B]                  &[$10^{-4}M_\odot$] &[$M_\odot$]   &[$10^{-3}M_\odot$] & &[$M_\odot$] \\
    \hline
    9(a)  &1.775  &0.111  &0.013\tianshu{(0.013)} &0.0019\tianshu{(0.0019)}     &0.061\tianshu{(0.061)}     &0.502&$4.88\times10^{-4}$ \\ 
    9(b)  &1.950  &0.094   &0.035\tianshu{(0.034)} &0.0063\tianshu{(0.0062)}     &0.065\tianshu{(0.064)}     &0.055&$3.35\times10^{-4}$\\ 
    9.25  &3.532  &0.124   &0.074\tianshu{(0.074)} &0.0106\tianshu{(0.0106)}     &0.116\tianshu{(0.116)}     &0.054&$3.17\times10^{-4}$\\ 
    9.5   &2.375  &0.142   &0.121\tianshu{(0.107)} &0.0160\tianshu{(0.0160)}     &0.180\tianshu{(0.165)}     &0.054&$2.24\times10^{-3}$\\ 
    11    &4.492  &0.321   &0.413\tianshu{(0.408)} &0.0418\tianshu{(0.0418)}     &0.763\tianshu{(0.759)}     &0.216&$3.02\times10^{-2}$\\ 
    15.01 &4.383  &0.288   &0.680\tianshu{(0.566)} &0.0548\tianshu{(0.0548)}     &1.212\tianshu{(0.931)}     &0.056&$5.64\times10^{-3}$\\
    16    &4.184  &0.390   &0.648\tianshu{(0.565)} &0.0653\tianshu{(0.0655)}     &1.025\tianshu{(0.850)}     &0.045&$1.24\times10^{-2}$\\
    17    &6.164  &1.123  &1.443\tianshu{(1.392)} &0.1123\tianshu{(0.1124)}     &2.706\tianshu{(2.611)}     &0.079 &$7.63\times10^{-3}$\\ 
    18    &8.617  &0.516   &1.055\tianshu{(1.032)} &0.1181\tianshu{(0.1182)}     &2.219\tianshu{(2.157)}     &0.054&$2.44\times10^{-2}$\\ 
    18.5  &6.408  &1.155  &1.192\tianshu{(1.064)} &0.1421\tianshu{(0.1426)}     &2.335\tianshu{(1.993)}     &0.043 &$8.96\times10^{-3}$\\
    19    &4.075  &0.466  &1.095\tianshu{(0.945)} &0.0949\tianshu{(0.0955)}     &1.515\tianshu{(1.188)}     &0.052&$7.06\times10^{-2}$ \\
    19.56$^*$ &3.890  &2.451                &1.619\tianshu{(1.284)} &0.2536\tianshu{(0.2550)}     &3.594\tianshu{(2.706)}     &0.047&0 \\
    20    &6.429  &0.881   &1.080\tianshu{(0.972)} &0.1087\tianshu{(0.1091)}     &2.126\tianshu{(1.832)}     &0.034&$1.28\times10^{-2}$\\ 
    23    &6.228  &0.513   &1.180\tianshu{(1.044)} &0.0944\tianshu{(0.0948)}     &1.627\tianshu{(1.346)}     &0.055&$1.92\times10^{-2}$\\ 
    24    &3.919  &0.779   &1.287\tianshu{(1.110)} &0.1348\tianshu{(0.1359)}     &2.155\tianshu{(1.525)}     &0.039&$2.64\times10^{-2}$\\
    25    &6.419  &1.101   &2.124\tianshu{(1.838)} &0.1887\tianshu{(0.1899)}     &3.128\tianshu{(2.491)}     &0.037&$3.18\times10^{-2}$\\ 
    40$^*$    &1.760  &1.804                &1.162\tianshu{(0.935)} &0.1798\tianshu{(0.1817)}     &4.661\tianshu{(3.661)}     &0.106&0 \\ 
    60    &8.026  &0.541   &1.645\tianshu{(1.572)} &0.1321\tianshu{(0.1323)}     &2.785\tianshu{(2.598)}     &0.117&$4.82\times10^{-2}$\\ 
    \end{tabular}
    \caption{This table summarizes some basic properties of our models. The core bounce time is set to $t=0$. $t_{\rm end}$ is the time after bounce when the full 3D simulation stops, $E_{\rm exp}$ is the explosion energy in Bethes ($\equiv$10$^{51}$ ergs). The isotopic yields are calculated at about twenty seconds post-bounce. \tianshu{$M^{\rm pow}$ is the mass calculated using the power-law tracer extrapolation method, while $M^{\rm exp}$ is calculated using the exponential tracer extrapolation method. Discussions and plots in this paper are based on the power-law results, while the exponential results are listed to indicate that there can be up to a $\sim$20\% uncertainty in certain isotope yields. A more detailed discussion on the uncertainties can be found in Section \ref{sec:compare}.} The solar stable Ni/Fe ratio is 0.058 \citep{lodders2021}, and the ratios in our models (including the residual progenitor envelope) range between about 0.6 to 7 times the solar value. $M_{\rm extra}$ is the mass of the neutrino-driven winds in the extrapolated phases. The 19.56 and 40 $M_\odot$ models form black holes at the end of the simulations \citep{burrows2023,burrows2024}; thus, they have $M_{\rm extra}=0$. }
    \label{tab:simulation_summary}
\end{table}

\begin{figure}
    \centering
    \includegraphics[width=0.96\textwidth]{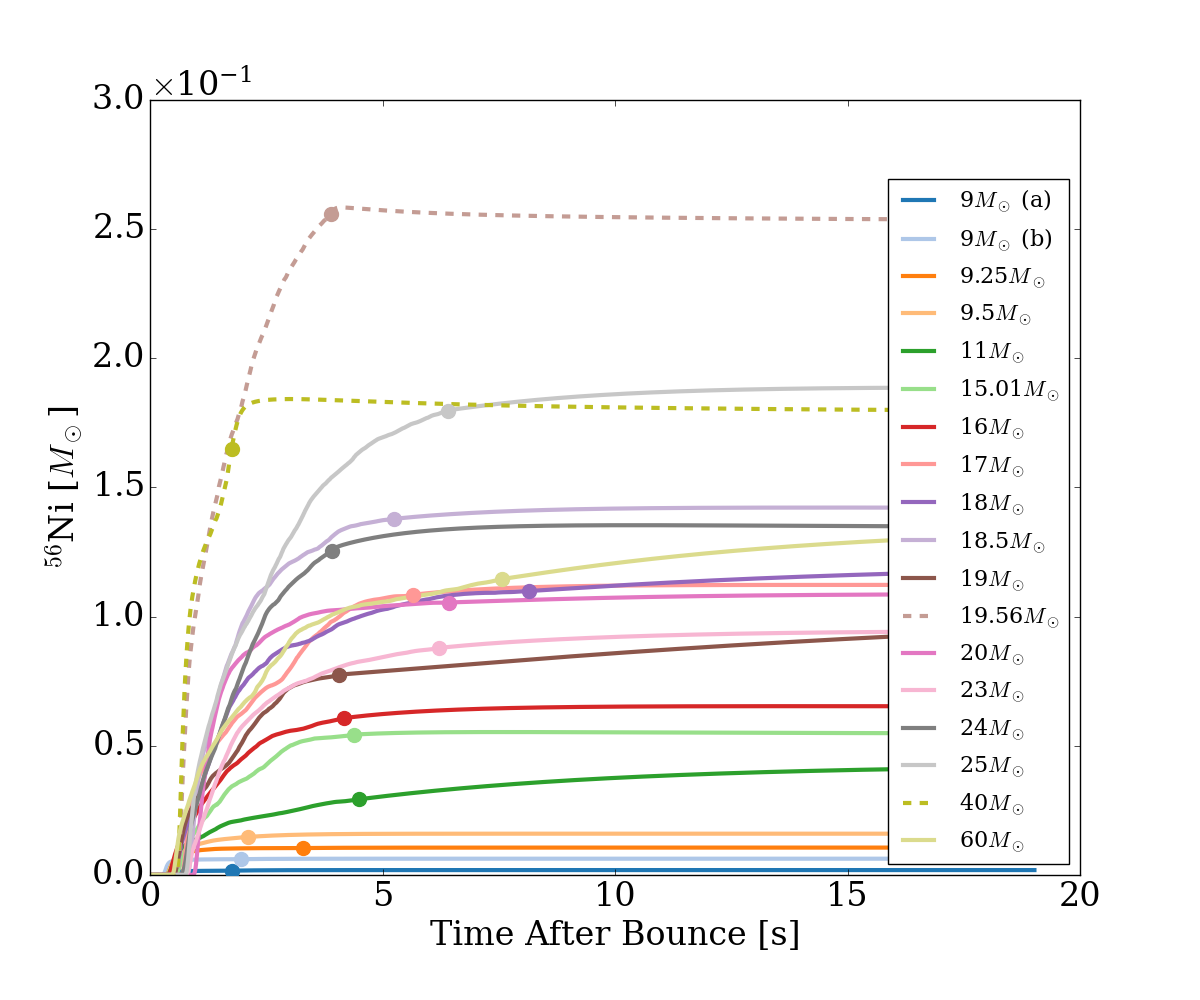}
    \caption{$^{56}$Ni yield as a function of time. The circular dot on each curve marks the beginning point of extrapolation. The black-hole formers (19.56 and 40 $M_{\odot}$) are plotted as dashed curves, and they have no extrapolated neutrino-driven winds. In all models, the majority of $^{56}$Ni is produced in the first 4$-$5 seconds after bounce, while the more aspherical models with stronger long-lasting accretion (e.g., the 11, 19, 25, and 60 $M_\odot$ models) experiences a further 10\% $^{56}$Ni production during the extrapolated phase. This late-time $^{56}$Ni increase results from the $\alpha$-rich freeze-out process in the enhanced neutrino-driven winds (see Figure \ref{fig:wind}) due to their longer-lasting accretion. However, in general after a few seconds the neutrino-driven winds don't contribute much to the total $^{56}$Ni yields.}
    \label{fig:ni56}
\end{figure}

\begin{figure}
    \centering
    \includegraphics[width=0.48\textwidth]{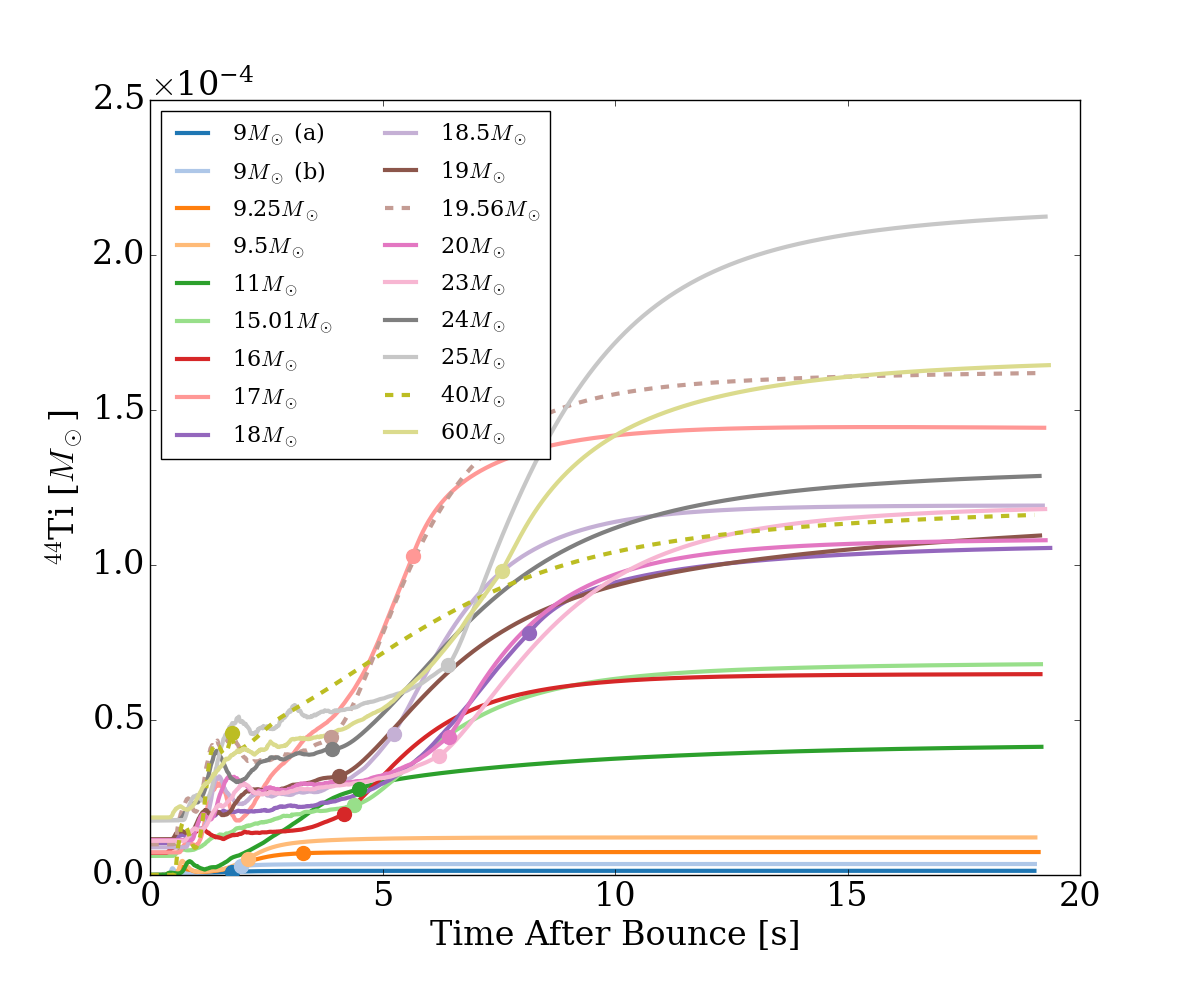}
    \includegraphics[width=0.48\textwidth]{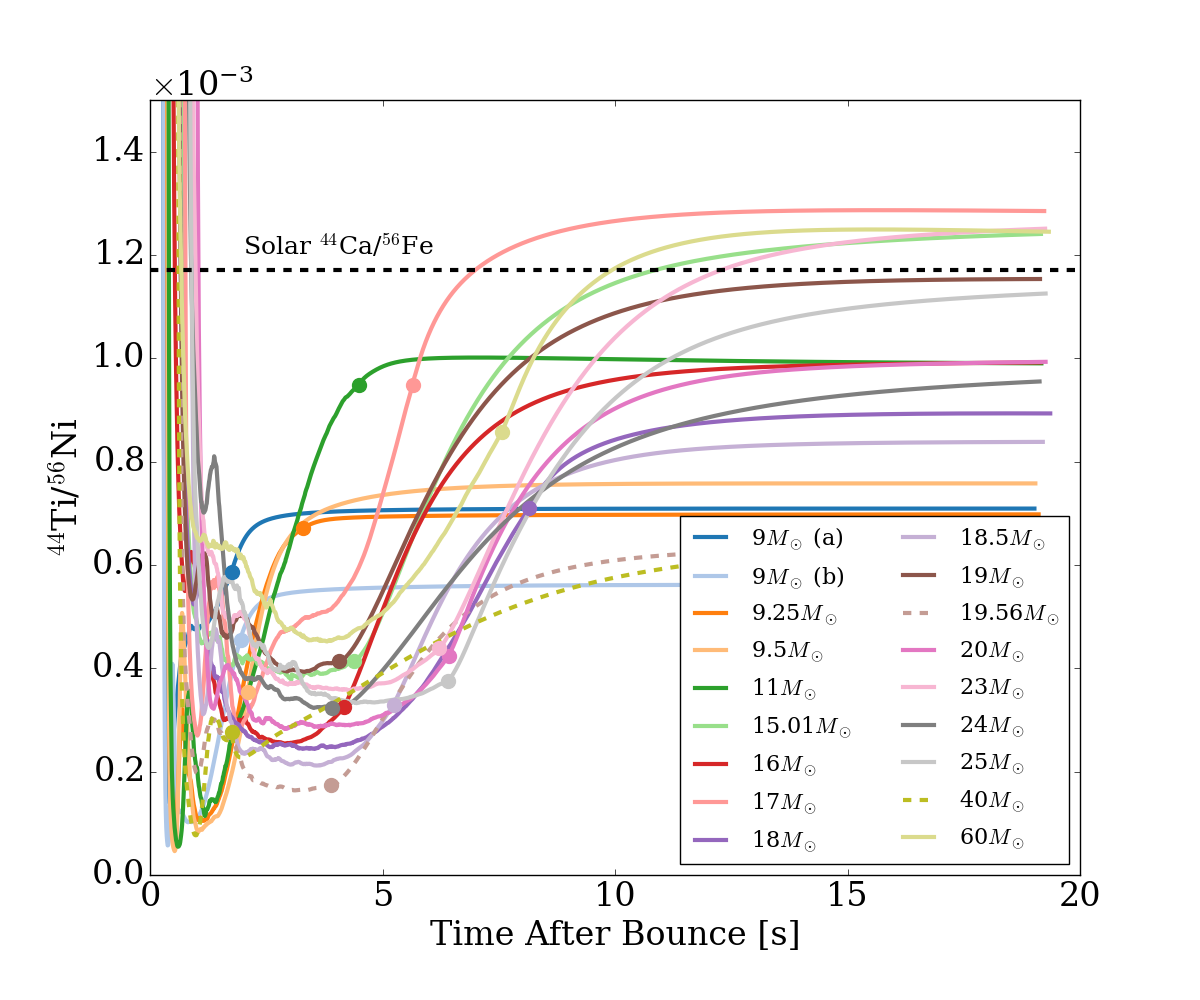}
    \caption{$^{44}$Ti yield (left) and $^{44}$Ti/$^{56}$Ni ratio (right) and as a function of time. The circular dot on each curve marks the beginning point of extrapolation. The black-hole formers (19.56 and 40 $M_{\odot}$) are plotted as dashed curves. The black horizontal line shows the solar $^{44}$Ca/$^{56}$Fe ratio as a reference. The asymptotic $^{44}$Ti yield is not achieved in most models until 10 seconds post bounce, except the lowest-mass ones (e.g. 9a, 9b, 9.25, and 9.5 $M_{\odot}$). The growth of the $^{44}$Ti yield in the extrapolated phase comes mostly from the $\alpha$-rich freeze-out process in the proton-rich neutrino-driven winds, which initially produces proton-rich isotopes such as $^{45}$V and $^{46}$Cr and then converts them into $^{44}$Ti as the temperature decreases. See text for a discussion.}
    \label{fig:ti44-ni56}
\end{figure}

\begin{figure}
    \centering
    \includegraphics[width=0.96\textwidth]{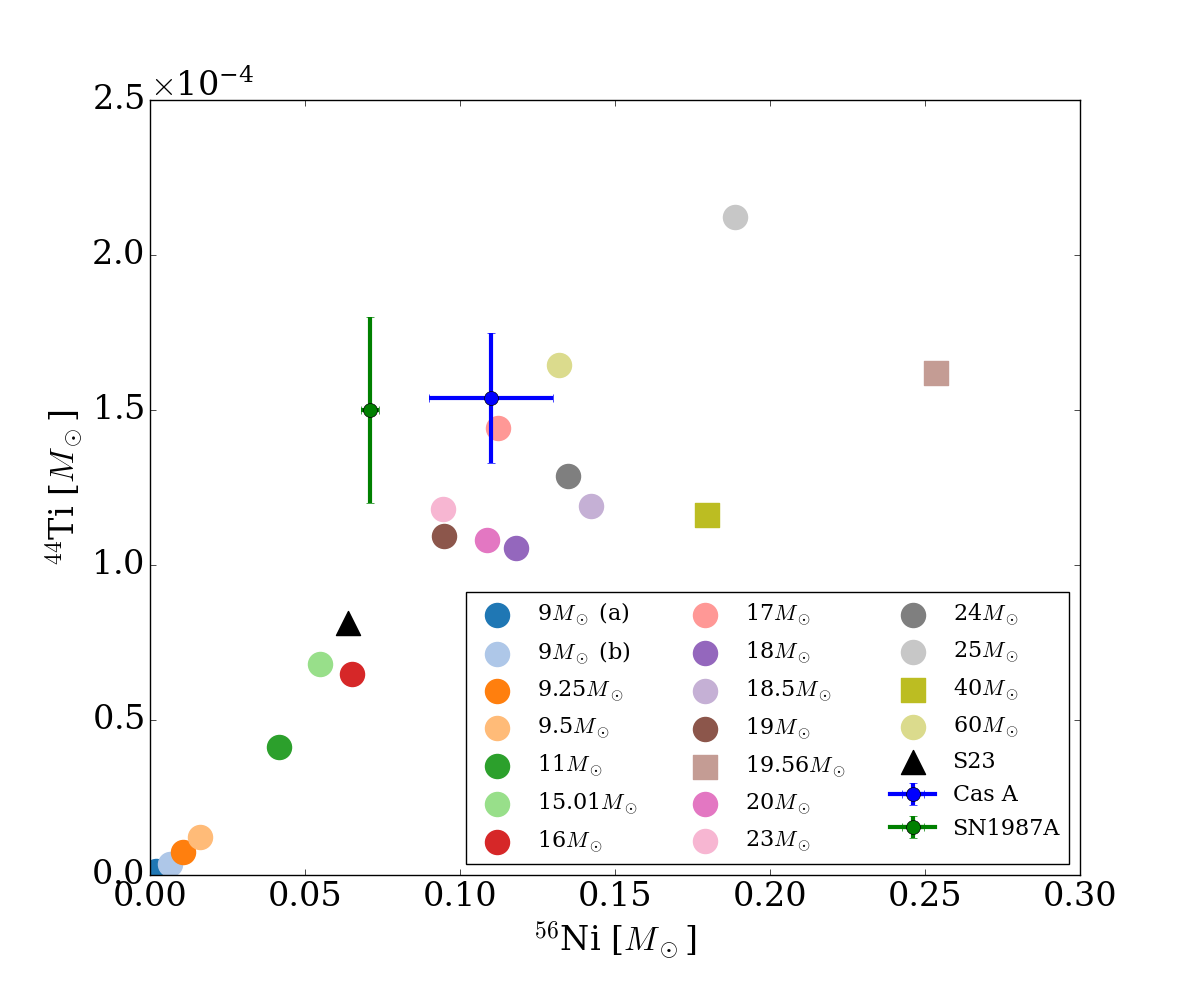}
    \caption{Final $^{44}$Ti and $^{56}$Ni yields of our theoretical 3D models compared to observed values in Cas A and SN1987A. The black-hole formers (19.56 and 40 $M_{\odot}$) are plotted as square dots. In addition to our 3D models, we include the 18.88 $M_\odot$ model from \citet{sieverding2023} (the black triangle, labeled as S23). The observed $^{44}$Ti and $^{56}$Ni masses of Cas A are taken from \citet{grefenstette2017} and \citet{hwang2012}, while the SN1987A values are taken from \citet{boggs2015} and \citet{mccray2016}. Our 3D models can easily produce the high amount of $^{44}$Ti and reach the high $^{44}$Ti/$^{56}$Ni ratio observed in Cas A. Therefore, the ``$^{44}$Ti-problem" \citep{the2006} is no longer an issue in self-consistent long-term, 3D, initially non-rotating CCSN models. In addition, the lower $^{44}$Ti/$^{56}$Ni ratios of the black-hole formers clearly show the important role of neutrino-driven winds in the production of $^{44}$Ti, since the winds in these models are turned off after black hole formation.}
    \label{fig:ti44-ni56-obs}
\end{figure}

\begin{figure}
    \centering
    \includegraphics[width=0.48\textwidth]{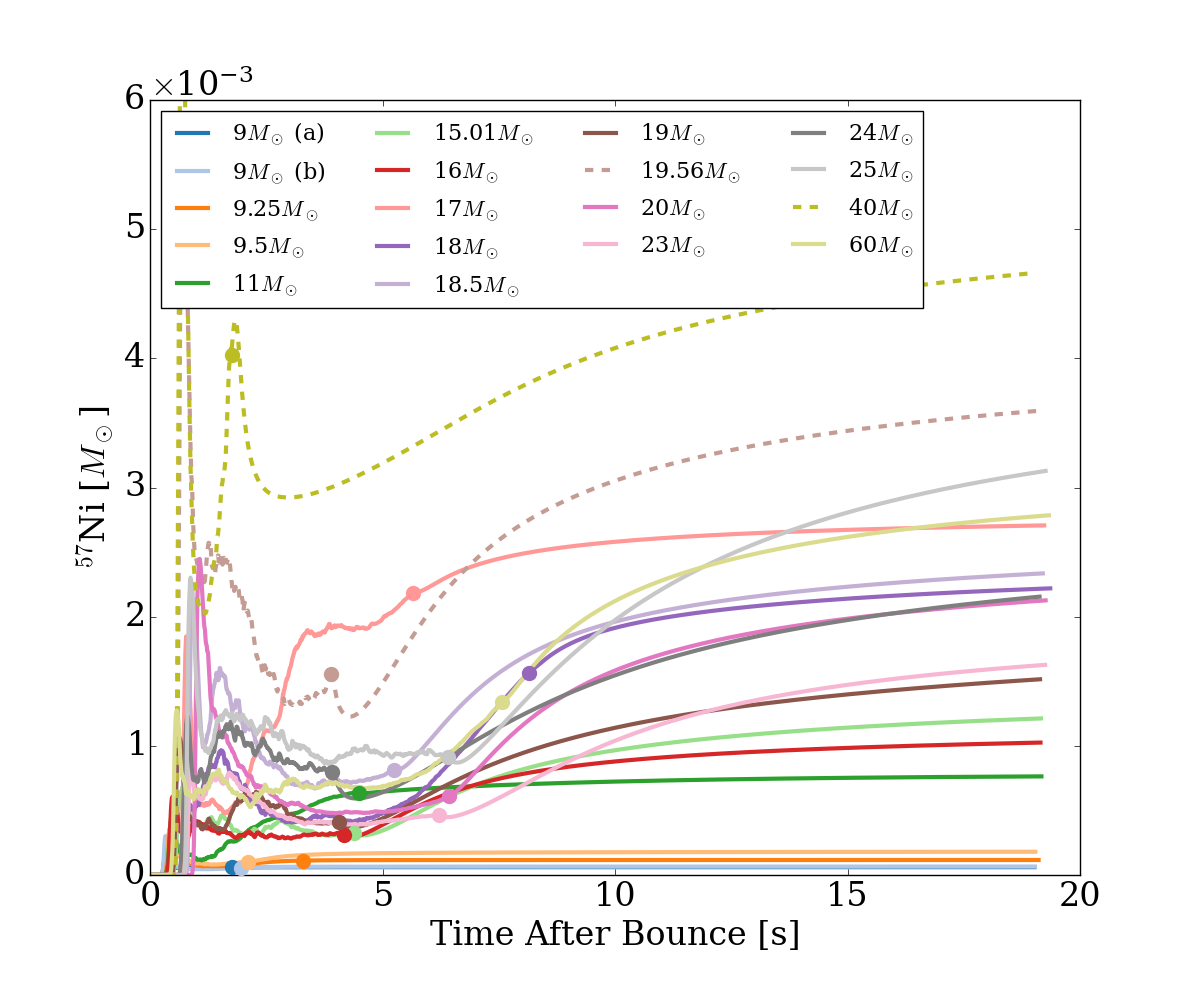}
    \includegraphics[width=0.48\textwidth]{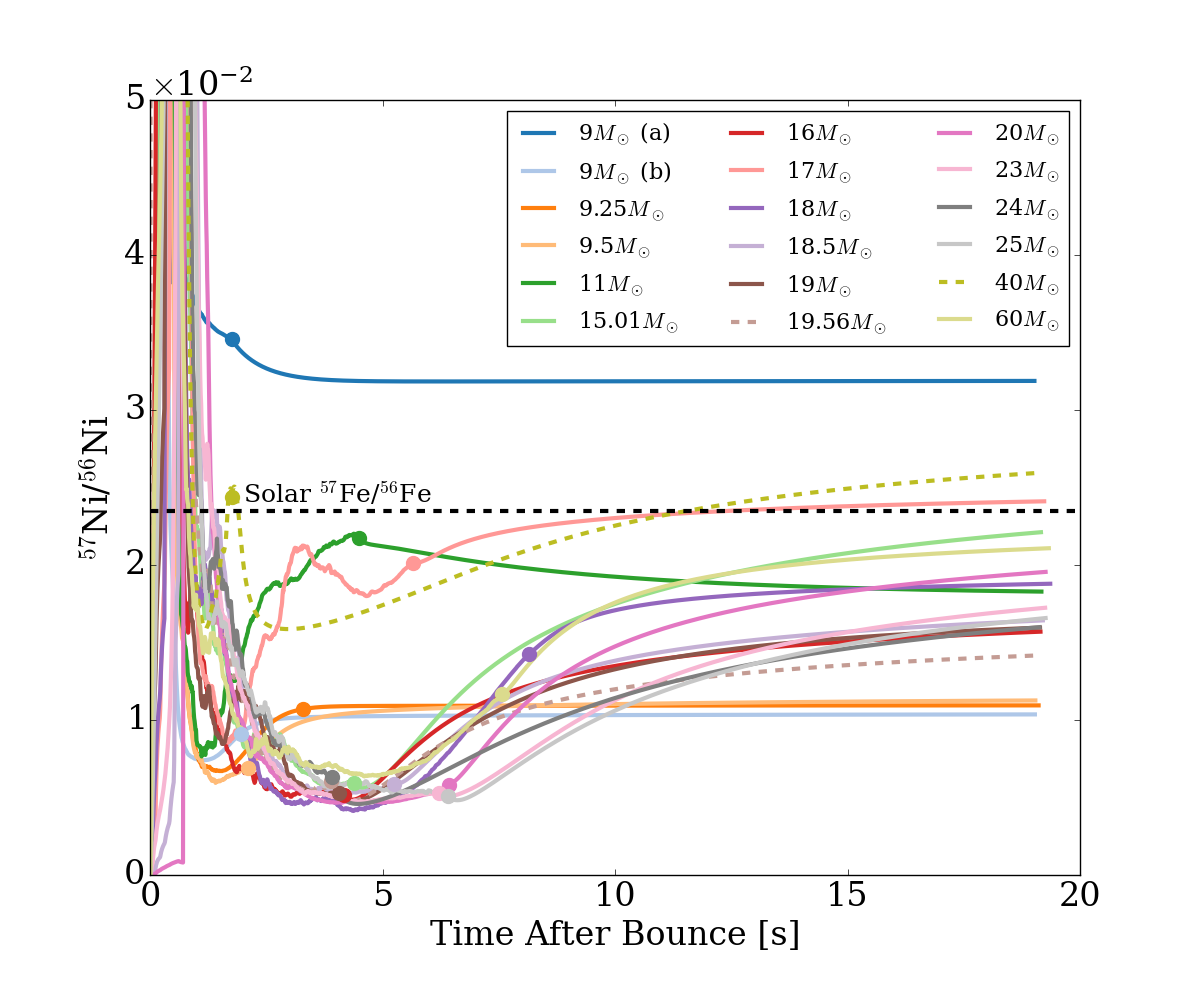}
    \caption{$^{57}$Ni yield and $^{57}$Ni/$^{56}$Ni ratio as a function of time. The circular dot on each curve marks the beginning point of extrapolation. The black-hole formers (19.56 and 40 $M_{\odot}$) are plotted as dashed curves. The black horizontal line shows the solar $^{57}$Fe/$^{56}$Fe ratio as a reference. Similar to $^{44}$Ti, a large amount of $^{57}$Ni is produced at later times ($>$5 seconds post-bounce) due to the $\alpha$-rich freeze-out process in the proton-rich neutrino-driven winds. However, a high $^{57}$Ni/$^{56}$Ni ratio (e.g., $3\times10^{-2}$) can't be reached unless the explosion has a significant amount of neutron-rich ejecta.}
    \label{fig:ni57-ni56}
\end{figure}

\begin{figure}
    \centering
    \includegraphics[width=0.96\textwidth]{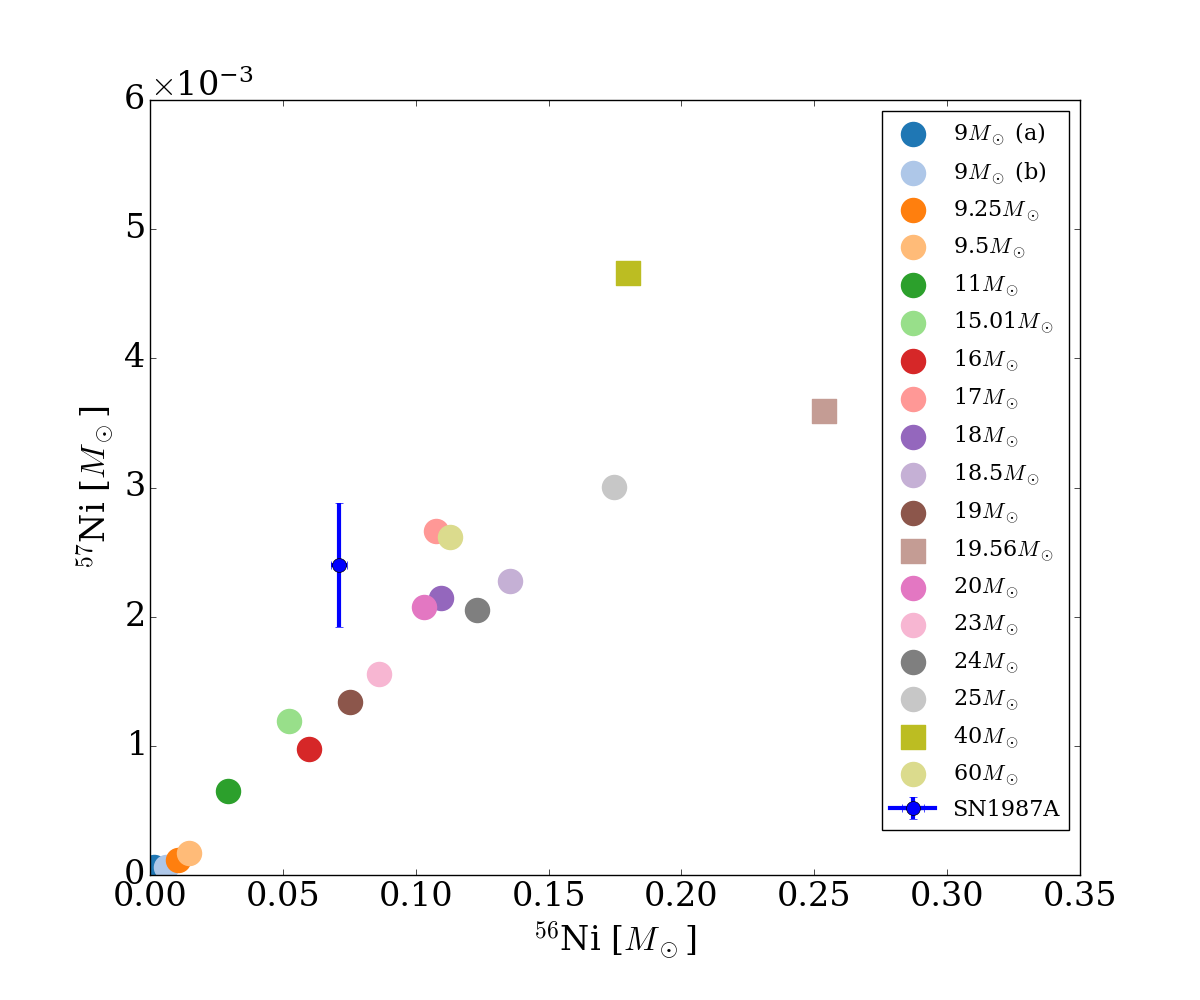}
    \caption{Final $^{57}$Ni and $^{56}$Ni yield of 3D theoretical models compared to observed values in SN1987A. The black hole formers (19.56 and 40) are plotted as square dots. Despite of the large error bars, SN1987A has a higher $^{57}$Ni/$^{56}$Ni ratio than most of our models. Only the most neutron-rich explosion (9a) shows a $^{57}$Ni/$^{56}$Ni ratio within the one-sigma range of the observation. One possibility is that SN1987A exploded more prompt than most of our massive models and ejected a larger amount of neutron-rich matter. See text for more discussions.}
    \label{fig:ni57-ni56-obs}
\end{figure}

\section{Results}\label{sec:results}
\subsection{$^{44}$Ti, $^{56}$Ni, and $^{57}$Ni}

$^{44}$Ti, $^{56}$Ni, and $^{57}$Ni are produced in CCSN explosions in measurable amounts and the radioactive decay of the latter two can help power supernova light curves. In addition, the presence and yields of these isotopes can be inferred by direct $\gamma$-ray and X-ray observations.

A range of observations have indicated that in both SN1987A and Cassiopeia A (Cas A) at least $10^{-4}M_\odot$ of $^{44}$Ti was produced, with $^{44}$Ti/$^{56}$Ni ratios above $10^{-3}$  \citep{hwang2004,hwang2012,grefenstette2014,boggs2015,mccray2016,grefenstette2017}. These values were inconsistent with previous theoretical CCSN models, in particular those that were spherically-symmetric  \citep{the2006,sukhbold2016,curtis2019,sieverding2023} and this $^{44}$Ti problem was identified almost two decades ago \citep{the2006}. One possible solution is the effect of explosion asymmetry. The strongly aspherical, though axisymmetric (2D), models of \citet{nagataki1997} and \citet{maeda2003} were able to obtain high ratios of $^{44}$Ti/$^{56}$Ni. However, these models used parameterized initial conditions and didn't explain the origin of the strong, even jet-like, asymmetries necessary. Such jet-like explosions are not commonly seen in self-consistent 3D simulations \citep{lentz2015,wongwathanarat2015,roberts2016,wongwathanarat2017,muller2017,oconnor2018,vartanyan2018,ott2018,summa2018,glas2019,burrows2019,nagakura2019,burrows2020,stockinger2020,bollig2021,sandoval2021,burrows2024}, unless the progenitor is initially rotating fast, which is thought to be a rare occurrence \citep{heger2005,obergaulinger2021,powell2023,shibagaki2024}. Three-dimensional simulations done by \citet{wongwathanarat2017} could explain the $^{44}$Ti production in Cas A, but those authors used prescribed neutrino luminosities at the inner grid boundary to tune the explosion energy to a desired value. The gray neutrino transport treatment they used was approximate, and their models could not produce enough $^{44}$Ti if the tracer electron fractions ($Y_e$s) were taken from the 3D simulations instead of by manually setting them to a value of 0.5. 

Recently, \citet{sieverding2023} published a self-consistent 3D initially non-rotating CCSN model that reached a $^{44}$Ti/$^{56}$Ni ratio close to observations. This paper emphasized the importance of long-term 3D simulations because (a) tracer trajectories in 3D show more complex temperature and density evolution and (b) the total ejecta mass continues to grow for many seconds. Both effects can aid in the production of $^{44}$Ti. Although this work is an important step in solving the $^{44}$Ti problem, the total amount of $^{44}$Ti in their model is still below $10^{-4}M_\odot$ and does not explain the observed values in Cas A and SN1987A. It was also unclear from their paper if the high $^{44}$Ti/$^{56}$Ni ratio is progenitor model dependent. Therefore, a systematic study of CCSN $^{44}$Ti production with many long-term 3D simulations has been lacking. In this section, we present for the first time such a systematic study of $^{44}$Ti production in initially non-rotating 3D CCSN explosion models, together with a corresponding discussion of $^{56}$Ni and $^{57}$Ni production and the effects of neutrino-driven winds. 

We start with a discussion of the production of $^{56}$Ni. In CCSN explosions, $^{56}$Ni is produced explosively by the shock wave as it traverses the mantle matter (predominantly the inner reaches of the oxygen shell) and in the neutrino-driven winds via $\alpha$-rich freeze-out \citep{woosley2002,nomoto2013,wanajo2023,arcones2023}. Explosive nucleosynthesis terminates quickly as the shock expands and the post-shock temperature drops, while the neutrino-driven winds can last many seconds post-bounce. \citet{wang2023} have shown that long-lasting accretion can strengthen the neutrino-driven winds by enhancing the neutrino luminosities and increasing the density in the wind formation region. Therefore, neutrino-driven winds play a more important role in nucleosynthesis in more aspherical 3D explosions with stronger long-lasting accretion (usually the case in more massive models) than in 1D spherical models.

Figure \ref{fig:ni56} shows the temporal evolution of $^{56}$Ni production in each simulation. The circular dot marks the point when model extrapolation begins. In all models, the vast bulk of $^{56}$Ni is produced in the first 4$-$5 seconds after bounce, meaning that the explosive nucleosynthesis component and the early-phase neutrino-driven winds dominate $^{56}$Ni production. More aspherical models with stronger long-lasting accretion (e.g., the 11, 19, 25, and 60 $M_\odot$ models) experience only $10-20$\% extra $^{56}$Ni production during the extrapolated phase. This late-time $^{56}$Ni growth comes from the $\alpha$-rich freeze-out process in the enhanced neutrino-driven winds due to long-lasting accretion (see Figure \ref{fig:wind}). The wind mass outflow rates of these models have shallower slopes and, thus, last for a longer period of time. However, in general after a few seconds the neutrino-driven winds don't contribute much to the total $^{56}$Ni yields. 

The production of $^{44}$Ti is different from that of $^{56}$Ni. \citet{magkotsios2011} has studied in detail the production channels of $^{44}$Ti, and we have used this work to identify the most important channel for $^{44}$Ti production. We find that the dominant contribution to $^{44}$Ti production in \tianshu{most of} our simulations\footnote{\tianshu{Model 9(a) is a special case because it has a strong neutron-rich phase. This phase suppresses the (p,$\gamma$)-leakage channel at early times and leads to a smaller amount of $^{44}$Ti. But because $^{56}$Ni is more suppressed, the $^{44}$Ti/$^{56}$Ni ratio in 9(a) is higher than that in the more proton-rich 9(b).}} comes from the neutrino-driven winds and  corresponds to the ``(p,$\gamma$)-leakage" regime discussed in \citet{magkotsios2011}. This is a special type of freeze-out. Briefly, in a proton-rich environment (which is very common in neutrino-driven winds), the (p,$\gamma$) reactions transfer the nucleosynthetic flow from symmetric to proton-rich isotopes. As the temperature decreases, the freeze-out process starts and the weak interactions decrease $Y_e$ and cause the flow to transfer back towards more stable isotopes. In the case of $^{44}$Ti, this means that isotopes such as $^{45}$V and $^{46}$Cr are produced at early times, and they are converted into $^{44}$Ti as the temperature decreases. This explains the non-monotonic growth of the $^{44}$Ti abundances in our tracers. The decay of $^{44}$V to $^{44}$Ti also contributes to $^{44}$Ti growth, but plays only a secondary role. The freeze-out process remains active even when the temperature is lower than 1 GK \tianshu{because of the ongoing weak interactions}. Therefore, the long production timescale of $^{44}$Ti \citep{sieverding2023,wang2024} is due to two main factors: (a) the neutrino-driven winds last for many seconds and keep contributing to the $^{44}$Ti yield and (b) the freeze-out process requires a very low temperature to reach the asymptotic state, and this takes a long time to achieve during the expansion. Figure \ref{fig:ti44-ni56} shows the $^{44}$Ti yields and $^{44}$Ti/$^{56}$Ni ratios of all our models as a function of time. The circular dot on each curve marks the point when the extrapolation is started. Many models take more than 10 seconds to reach their asymptotic $^{44}$Ti yields, and $10^{-4}M_\odot$ of $^{44}$Ti production can easily be achieved by models more massive than $\sim$16 $M_\odot$. More massive models in general produce more $^{44}$Ti (see Table \ref{tab:simulation_summary}). The right panel shows that the more massive models also tend to have higher $^{44}$Ti/$^{56}$Ni ratios than the low mass models (e.g., 9a, 9b, 9.25, and 9.5 $M_{\odot}$), so the more massive models not only have higher ejecta masses but also a higher $^{44}$Ti production efficiency. This is because more massive explosion models are in general more asymmetric than the low-mass progenitor models due to their generally more massive envelopes. 

\tianshu{Explosion asymmetry and general multi-dimensional effects play} central roles in the production of $^{44}$Ti. As \citet{sieverding2023} has pointed out, large-scale convection leads to non-monotonic thermal histories. Thus, nucleosynthesis can experience multiple nucleosynthetic phases, which aid the production of $^{44}$Ti. We see the same behavior in our 3D simulations. However, this effect can't solely explain the significantly higher $^{44}$Ti yields in our simulations, since the ejecta with significantly non-monotonic thermal histories contribute only a small fraction to the $^{44}$Ti yields. The most important 3D effect is the so-called ``simultaneous explosion and accretion" phenomenon. This phenomenon is sometimes referred to as ``long-lasting" or ``fallback" accretion. In the context of this phenomenon, the stellar envelope above the proto-neutron star (PNS) is not immediately swept away by the shock wave after the launch of the explosion. Instead, it forms a few inward-directed funnels that extend almost to the PNS surface. Matter in the outer layers can fall very close to the PNS through these funnels, after which it is heated by neutrinos and ejected as part of the anisotropic neutrino-driven wind. The simultaneous explosion and accretion phenomenon significantly increases the amount of mass interacting with neutrinos, which adds to the $\alpha$-rich freeze-out component. This enhancement of $\alpha$-rich freeze-out lasts for many seconds, until accretion is completely terminated and a spherical wind develops\footnote{For examples of spherical winds see \citet{stockinger2020} and \citet{wang2024b}}. In addition, interactions with neutrinos push $Y_e$ to higher values, which provides the proton-rich environment required by the ``(p,$\gamma$)-leakage" channel. Therefore, the simultaneous explosion and accretion phenomenon (redirecting fallback into a component of the late-time winds) aids the production of $^{44}$Ti by enhancing the proton-rich neutrino-driven winds. This is the origin of the high $^{44}$Ti yields we observed in our long-term 3D CCSN simulations. \tianshu{Hence, more asymmetric explosions naturally lead to more vigorous simultaneous explosion and accretion behavior and, thus, to higher $^{44}$Ti production.}

Neutrino-driven winds are important for the production of $^{44}$Ti because they are the major context of $\alpha$-rich freeze-out. Thus, it is crucial to capture the behavior of neutrino-driven winds with tracers. For post-processed tracers, the backward integration method has two major advantages: (a) this method inserts the tracers at the end of the simulations, so all ejecta regions can be identified and represented and (b) this method reproduces the freeze-out conditions more accurately than the forward method, since it does not need to follow the tracers through the complicated fluid motions in the neutrino-driven convection region \citep{reichert2023,sieverding2023b,wang2023}. Therefore, the backward method gives more accurate nucleosynthetic results than the forward method. 

Figure \ref{fig:ti44-ni56-obs} plots the $^{44}$Ti and $^{56}$Ni yields predicted by our 3D theoretical CCSN models, along with the observed Cas A and SN1987A yields. The observed $^{44}$Ti and $^{56}$Ni masses of Cas A are taken from \citet{grefenstette2017} and \citet{hwang2012}, while the SN1987A values are taken from \citet{boggs2015} and \citet{mccray2016}. Our 3D models can easily produce high amounts of $^{44}$Ti and reach the high $^{44}$Ti/$^{56}$Ni ratio observed in Cas A. \tianshu{The possible presence of unshocked Fe-rich material in Cas A may move the data point rightward, which would be even more consistent with our predictions.} In addition, the lower $^{44}$Ti/$^{56}$Ni ratios of the black-hole formers clearly show the important role of neutrino-driven winds in the production of $^{44}$Ti, since the winds in these models are turned off after the black hole formation. Therefore, the ``$^{44}$Ti-problem" \citep{the2006} is no longer an issue in self-consistent long-term 3D initially non-rotating CCSN models. SN1987A is still about two sigmas above the $^{44}$Ti/$^{56}$Ni ratios predicted by our models. We now turn to a discussion of this issue and its corresponding $^{57}$Ni/$^{56}$Ni ratio.

$^{57}$Ni is an interesting radioactive isotope. Figure \ref{fig:ni57-ni56} shows the temporal evolution of the $^{57}$Ni yields and $^{57}$Ni/$^{56}$Ni ratios in our models. Two classes of behaviors are seen. If the explosion has ever experienced a neutron-rich phase (e.g., models 9a, 11, and 17), the $^{57}$Ni/$^{56}$Ni ratio quickly rises to a high value during the neutron-rich phase and stays there. Although  $^{56}$Ni production in the neutrino-driven winds may decrease this ratio a bit, the overall $^{57}$Ni/$^{56}$Ni ratios of these models remain relatively high. For models with no neutron-rich ejecta, the production timescale of $^{57}$Ni is even longer than that of $^{44}$Ti. This is because $^{57}$Ni is also produced in significant amounts in the $\alpha$-rich freeze-out process in proton-rich neutrino-driven winds. The asymptotic $^{57}$Ni/$^{56}$Ni ratios in these two different classes are not as different as they are initially, unless the neutron-rich phase occurs so early that the explosive nucleosynthesis component (which has more mass than the winds) partly becomes neutron-rich (witness 9a)\footnote{The ejecta $Y_e$ distribution for all models can be found in \citet{wang2024}}.

Figure \ref{fig:ni57-ni56-obs} shows the $^{57}$Ni and $^{56}$Ni yields predicted by 3D theoretical CCSN models and observed in SN1987A. The $^{57}$Ni amount of SN1987A is taken from \citet{kurfess1992} and the $^{56}$Ni amount is taken from \citet{hwang2012}. Similar to the $^{44}$Ti case, SN1987A has a $^{57}$Ni/$^{56}$Ni ratio about 1.5 sigmas above the values seen in our 3D simulations. One possible explanation for the higher $^{44}$Ti/$^{56}$Ni and $^{57}$Ni/$^{56}$Ni ratios in SN1987A is that the explosion might have had an early and strong neutron-rich phase, so that a fraction of the matter participating in explosive nucleosynthesis becomes neutron-rich. This will enhance $^{57}$Ni production, but the most significant influence is that the production of $^{56}$Ni at the early phase will be suppressed. $^{44}$Ti is not influenced by this early neutron-rich phase since it is produced mostly in the wind phase. Therefore, a more neutron-rich explosion, which might be caused by a more prompt explosion, will naturally lead to higher $^{44}$Ti/$^{56}$Ni and $^{57}$Ni/$^{56}$Ni ratios. This is one possibility for SN1987A.

As the name indicates, $^{4}$He can be produced in significant amounts by the $\alpha$-rich freeze-out. Therefore, $^{4}$He exists both in the helium envelope and in the innermost ejecta where the $^{56}$Ni abundance is high. The coexistence of $^{4}$He with $^{56}$Ni is interesting because it can absorb the $\gamma$-rays emitted by the decay of $^{56}$Ni and influence the emergent spectra. $^{4}$He in the nickel/iron-rich region of supernova remnants might be a potentially interesting observable. We find that the $^{4}$He/$^{56}$Ni ratios in the $^{56}$Ni-rich ejecta ($X_{^{56}\rm Ni}>10^{-5}$) are around 0.4$-$0.5 in most our models, while the low-mass models (9a, 9b, 9.25, 9.5, and 11 $M_{\odot}$) have higher $^{4}$He/$^{56}$Ni ratios around 0.8 \footnote{The $^{4}$He/$^{56}$Ni ratios in the $^{56}$Ni-rich ejecta of all our models are: (9a: 0.794), (9b: 0.893), (9.25: 0.912), (9.5: 0.786), (11: 0.810), (15.01: 0.440), (16: 0.474), (17: 0.638), (18: 0.416), (18.5: 0.426), (19: 0.514), (19.56: 0.388), (20: 0.507), (23: 0.528), (24: 0.402), (25: 0.466), (40: 0.375), (60: 0.491).}.

\subsection{Stable Isotopes of Nickel and Stable Nickel to Iron Ratios}

Not only does the expansion timescale influence the electron fraction ($Y_e$) during the early explosion phase, interactions between neutrino-driven winds and fallback accretion can change the outflow velocities and influence the wind $Y_e$s. In some models (11, 17, 40 and 60 $M_{\odot}$), we see neutron-rich episodes during the neutrino-driven wind phase. The mass ejected during these periods is very small compared to the mass ejecta during the early explosion phases and the ejecta entropy is higher. These neutron-rich phases can easily produce heavy elements up to $^{90}$Zr \citep{wang2023,wang2024}. One observable in supernova remnants, the stable Ni/Fe ratio\footnote{The stable Ni and Fe isotopes are $^{58}$Ni, $^{60}$Ni, $^{61}$Ni, $^{62}$Ni, $^{64}$Ni, $^{54}$Fe, $^{56}$Fe, $^{57}$Fe, and $^{58}$Fe. The ratios we provide include the entire progenitor envelope mass at the time of collapse, according to the progenitor evolutions conducted by \citet{sukhbold2016,sukhbold2018}.}, is dependent upon the strength of such neutron-rich phases. There are roughly three classes of models in our set. The first class contains the most neutron-rich models in which the early explosion phase has $Y_e<0.5$ (e.g., model 9a, and the models in \citet{wang2024b}). These models have Ni/Fe ratios more than 7 times as high as the solar value, and are close to the recent observation of the Crab nebula \citep{tea2024}. The second class contains the models with neutron-rich episodes in their winds (11, 17, 40, and 60 $M_{\odot}$). The Ni/Fe ratios in these models are about 1.5 to 5 times as high as the solar value, depending upon the amount of mass ejected during the neutron-rich period. The last class represents all models without any neutron-rich period. These models all have sub-solar Ni/Fe ratios (as low as 0.6 times the solar value). Although the neutron-rich wind phases play an important role in the production of many interesting isotopes, their appearance depends upon many aspects of the explosion dynamics and is very difficult to model or discern in advance of simulation. We list the asymptotic ratios of stable nickel to stable iron that we find for this suite of 18 long-term 3D CCSN models in Table \ref{tab:simulation_summary}. The shape and direction of the wind regions and accretion funnels, the PNS properties, and the outer envelope structures may all have an impact on the wind electron fraction. Therefore, we can conclude only that the neutrino-driven winds are mostly proton-rich, but with episodic neutron-rich episodes.


\subsection{Effects of the Extrapolation Method}
\label{sec:compare}

Since a significant amount of $^{44}$Ti and $^{57}$Ni is produced during the extrapolated phase, will the above results be changed by the choice of extrapolation method? The contribution from the extrapolated neutrino-driven winds accounts for no more than 20\% of the yield of these radioactive isotopes\footnote{\tianshu{Neutrino-driven winds start shortly after the explosion \citep{wang2023}, and the first few seconds of wind evolution before the extrapolated phase have been captured by our 3D simulations. Hence, since the wind mass outflow rates decay approximately exponentially with time (Figure \ref{fig:wind}), the bulk of the wind mass ejection occurs before extrapolation. What is left is the long-tail contribution from neutrino-driven winds during the extrapolated phase, whose estimated masses are listed as $M_{\rm extra}$ in Table \ref{tab:simulation_summary}. The contribution to each individual isotope is then calculated by multiplying $M_{\rm extra}$ with the corresponding isotopic mass fractions at late time: $^{56}$Ni$\sim$0.45, $^{44}$Ti$\sim$0.0004, and $^{57}$Ni$\sim$0.004. Given the small $M_{\rm extra}$, the extrapolated winds can not have a strong impact on the final nucleosynthetic results.}}, so the wind extrapolation does not influence the results to any significant degree. However, it is not impossible that how we chose to extrapolate the tracer thermal histories might change our conclusions, since this influences most of the ejecta that never reaches high temperatures. Therefore, it would be helpful to have a robustness test of our choice of tracer extrapolation method.

In this subsection, we explore the alternate (exponential) tracer extrapolation method:
\begin{equation}
    \begin{split}
        T(t) =& T(t_{\rm end})\exp(-\frac{t-t_{\rm end}}{3\tau}),\\ 
        \rho(t) =& \rho(t_{\rm end})\exp(-\frac{t-t_{\rm end}}{\tau}),\\ 
    \end{split}
\end{equation}
where $t_{\rm end}$ is the post-bounce duration of the 3D simulations and $\tau$ is an expansion timescale fitted using the last 10 ms of the trajectories. 
The exponential method leads to a faster decrease of temperature and density than the power-law method. Therefore, nucleosynthesis will terminate significantly earlier if the exponential method is used. 

\tianshu{The results for the exponential method applied to all models are listed in Table \ref{tab:simulation_summary}. The uncertainty of the $^{56}$Ni yield due to extrapolation method is $<$1\%, while it's $\sim$10\% for $^{44}$Ti and $\sim$20\% for $^{57}$Ni.} \tianshu{The different sensitivities to the extrapolation method for these three isotopes are directly related to their different production timescales. Isotopes with longer production timescales will be more dependent upon the choice of extrapolation method.} A $\sim$20\% uncertainty is not negligible, but it is tolerable in the context of current CCSN nucleosynthesis studies and does not change our general conclusions. The impact of the choice of extrapolation method is weaker in some models(e.g. the 9.x models, 17, 18, and 60 models), as they have already captured much of the late-time yield growth.  

A more realistic tracer evolution at late times might be a combination of the power-law and exponential relations. When the tracer is ejected from the surface of the PNS in the neutrino-driven wind, it achieves a radial velocity of a few tens of thousands of kilometers per second. This high velocity leads to a fast temperature and density decrease, with a slope similar to that of the exponential method. When the tracer reaches the wind termination shock where the neutrino-driven wind catches up with the relatively slowly-moving shocked ejecta, its temperature and density evolution switch to the second mode after a short period of shock heating. This second phase has a slope more similar to that obtained using the power-law method. Therefore, an extrapolation method similar to the one employed in \citet{harris2017} and \citet{sieverding2023} might be a better choice, which concatenates the exponential and power-law methods. However, as the position of the wind termination shock is changing, the temperature at which the two methods should be linked is also time-varying. In addition, the wind region can be significantly deformed from a sphere due to interactions with infalling matter \citep{wang2023}, so there can be a wide range of wind termination shock radii and temperatures, even in a single snapshot of a simulation. Therefore, the only way to unambiguously calculate the late-time evolution of CCSN explosions is to carry the 3D radiation-hydro simulations to 15$-$20 seconds post-bounce.


\section{Conclusion}\label{sec:conclusion}

This paper presents the late-time evolution of the yields of $^{44}$Ti, $^{56}$Ni, $^{57}$Ni, and stable iron and nickel isotopes for 18 3D CCSN simulations. The ejecta isotopic abundances at about twenty seconds post-bounce are calculated, the last phases by an extrapolation method of both tracer thermal histories and neutrino-driven wind mass outflow rates from our already long-term 3D simulations. For the first time, a detailed systematic study reveals the theoretical asymptotic yields of these interesting nucleosynthetic products, informed by state-of-the-art long-term 3D CCSN simulations.

In our models, the majority of $^{56}$Ni is produced in the first 4$-$5 seconds after bounce, revealing that the explosive nucleosynthetic component and the early-phase neutrino-driven wind dominate $^{56}$Ni production. A clear relation between final $^{56}$Ni yield and explosion energy has previously been shown in \citet{wang2024} and \citet{burrows2024}.

We find that the production timescales of $^{44}$Ti are significantly longer than those of $^{56}$Ni. Some models take more than 15 seconds to reach their asymptotic $^{44}$Ti abundances. We find that $^{44}$Ti is mostly produced in neutrino-driven winds via the ``(p,$\gamma$)-leakage" channel \citep{magkotsios2011}. In this special proton-rich variant of freeze-out, the (p,$\gamma$) reactions transfer the nucleosynthetic flow from symmetric to proton-rich isotopes. As the temperature decreases, the freeze-out process starts and weak interactions decrease $Y_e$ and cause the flow to revert to more stable isotopes. In this process, isotopes such as $^{45}$V and $^{46}$Cr are produced at early times, and they are slowly converted back to $^{44}$Ti as the temperature decreases. This freeze-out process remains active even when the temperature is lower than 1 GK. Therefore, there are two main reasons for the long production timescale of $^{44}$Ti. First, the neutrino-driven wind phase in which $^{44}$Ti is produced lasts for many seconds post-bounce. The simultaneous explosion and accretion phenomenon in 3D CCSN simulations allows the outer stellar matter to fall very close to the PNS through accretion funnels, which significantly increases the amount of mass interacting with neutrinos and leads to extended and stronger neutrino-driven winds \citep{wang2023}. This enhancement of wind strength lasts for many seconds until accretion is completely terminated and a spherical wind develops \citep{stockinger2020,wang2024b}. Second, the freeze-out process requires a very low temperature (below 1 GK) to reach the asymptotic abundances, which also takes many seconds to achieve. 

Our 3D models produce $^{44}$Ti ranging from $10^{-6}M_\odot$ to $2\times10^{-4}M_\odot$. The high $^{44}$Ti/$^{56}$Ni ratio observed in Cas A can be achieved by \tianshu{some} of our models\tianshu{, though the data point is still on the high side}. \tianshu{Given the fact that Cas A may still contain some unshocked Fe-rich matter, the actual $^{44}$Ti/$^{56}$Ni ratio might be a bit lower, which would then be even more consistent with our models}. Therefore, the ``$^{44}$Ti-problem" \citep{the2006} is no longer an issue in self-consistent long-term, 3D, initially non-rotating CCSN models. 

We find that there are various pathways to $^{57}$Ni production. If the explosion has ever experienced a neutron-rich phase, the $^{57}$Ni/$^{56}$Ni ratio quickly rises to a high value during the neutron-rich phase and stays there or decreases slowly due to $^{56}$Ni production in the neutrino-driven wind. For models with no neutron-rich ejecta, the production timescale of $^{57}$Ni is even longer than that of $^{44}$Ti. This is because $^{57}$Ni is also produced in significant amounts in the $\alpha$-rich freeze-out process in proton-rich neutrino-driven winds. The asymptotic $^{57}$Ni/$^{56}$Ni ratios for the two pathways are not as different as they are initially, unless the neutron-rich phase is early and strong.

The observed $^{44}$Ti/$^{56}$Ni and $^{57}$Ni/$^{56}$Ni ratios in SN1987A are both $\sim$2 sigma above the values seen in our 3D simulations. One possible explanation is that the actual explosion might have an early and strong neutron-rich phase, so that a fraction of the explosive nucleosynthetic products become neutron-rich. This will enhance $^{57}$Ni production, but the most significant influence is that the production of $^{56}$Ni at the early phase will be suppressed. $^{44}$Ti is not influenced by this early neutron-rich phase, since it is produced mostly in the wind phase. Therefore, a more neutron-rich explosion, which might be caused by a more prompt explosion, will naturally lead to higher $^{44}$Ti/$^{56}$Ni and $^{57}$Ni/$^{56}$Ni ratios. 

To test the robustness of the above findings, we changed our tracer extrapolation method from a power-law method to an exponential method. We find that the choice of extrapolation method changes the final yields by no more than 20\%, even in our worst case. This is not a negligible influence, but a $\sim$20\% uncertainty is tolerable for current CCSN nucleosynthesis studies and it will not change our general and substantive conclusions. However, the only way robustly to determine the late-time nucleosynthetic yields of core-collapse supernovae is to carry out detailed 3D supernova simulations to 15$-$20 seconds post-bounce. This project, though daunting, is now almost within reach of modern codes and machines, but will require a concerted effort to realize.


We also briefly discuss the stable Ni/Fe ratios. Our models have stable Ni/Fe ratios for all the ejecta of the explosion ranging from sub-solar values (as low as 0.6 times the solar value) to seven times the solar values. The latter is similar to the recent observation value of the Crab nebula \citep{tea2024}. However, Ni/Fe ratios, and in general the production of neutron-rich isotopes, depend strongly upon the strength of neutron-rich phases in the explosions. Since such phases are sensitive to many aspects of explosion dynamics, their occurrence seems to be stochastic. Therefore, the uncertainties of neutron-rich isotope abundances can be very large.

It is important to list some limitations of this work. First, our results rely on extrapolations of tracer thermal histories and neutrino-driven wind outflow rates. The accuracy of these extrapolations has not been tested by 3D simulations carried out to the same timescale. Second, although this is a systematic study covering a number of different progenitor models, all progenitors are calculated using the code Kepler \citep{sukhbold2016,sukhbold2018} and it is unclear if these progenitors cover the same range of structures and properties of real stars. In addition, the nuclear reaction rates are all taken from the JINA Reaclib \citep{cyburt2010}, for which there are still large uncertainties in some rates \citep{bliss2020}. Although we have mentioned briefly the potential roles of initial perturbations and rotation, a detailed study of their effects is still lacking. There are also uncertainties in the nuclear equation of state and the neutrino opacities/emissivities, etc.

\section*{Acknowledgments}
We acknowledge our ongoing and fruitful collaborations with David Vartanyan and Christopher White. We also acknowledge support from the U.~S.\ Department of Energy Office of Science and the Office of Advanced Scientific Computing Research via the Scientific Discovery through Advanced Computing (SciDAC4) program and Grant DE-SC0018297 (subaward 00009650), support from the U.~S.\ National Science Foundation (NSF) under Grants AST-1714267 and PHY-1804048 (the latter via the Max-Planck/Princeton Center (MPPC) for Plasma Physics), and support from NASA under award JWST-GO-01947.011-A.  A generous award of computer time was provided by the INCITE program, using resources of the Argonne Leadership Computing Facility, a DOE Office of Science User Facility supported under Contract DE-AC02-06CH11357. We also acknowledge access to the Frontera cluster (under awards AST20020 and AST21003); this research is part of the Frontera computing project at the Texas Advanced Computing Center \citep{stanzione2020} under NSF award OAC-1818253. Finally, the authors acknowledge computational resources provided by the high-performance computer center at Princeton University, which is jointly supported by the Princeton Institute for Computational Science and Engineering (PICSciE) and the Princeton University Office of Information Technology, and our continuing allocation at the National Energy Research Scientific Computing Center (NERSC), which is supported by the Office of Science of the U.~S.\ Department of Energy under contract DE-AC03-76SF00098.

\section*{Data Availability}
This paper uses previously published data, which can be accessed via\dataset[DOI: 10.5281/zenodo.10498614]{https://doi.org/10.5281/zenodo.10498614} \citep{wang_2024_10498614}. This online repository includes all the tracer trajectories (masses, positions, temperatures, electron fractions, and local neutrino spectra), together with the ejecta abundances of each model at the end of the 3D simulations. The progenitor models from \citet{sukhbold2016} and \citet{sukhbold2018} are available as well. The 17, 18, 18.5, 20, 25, and 60 $M_\odot$ models have been evolved to later times in 3D since the publication of the above repository. However, we can not replace the files in the repository, as this would significantly exceed to size limit of the repository. Therefore, we encourage readers to start with the online repository, and if necessary, to make direct requests to the authors for the rest of data.

\bibliography{sample631}{}
\bibliographystyle{aasjournal}




\end{document}